\documentclass[journal]{IEEEtran}
\usepackage[utf8]{inputenc}
\usepackage{graphicx,cite}
\usepackage{amsmath,amssymb}
\usepackage{flushend}
\usepackage{algorithm,algorithmic,multicol}

\author{Evangelos Vlachos, George~C.~Alexandropoulos,~\IEEEmembership{Senior~Member,~IEEE} and John Thompson,~\IEEEmembership{Fellow,~IEEE}
\thanks{E. Vlachos and J. Thompson are with Institute for Digital Communications, University of Edinburgh, Edinburgh, EH9 3JL, UK (e-mails: \{e.vlachos, j.s.thompson\}@ed.ac.uk).}
\thanks{G. C. Alexandropoulos is with the Mathematical and Algorithmic Sciences Lab, Paris Research Center, Huawei Technologies  France SASU, 92100 Boulogne-Billancourt, France (e-mail: george.alexandropoulos@huawei.com).
}}

\begin{document}

\title{Massive MIMO Channel Estimation for\\ Millimeter Wave Systems via Matrix Completion}

\maketitle

\begin{abstract}
Millimeter Wave (mmWave) massive Multiple Input Multiple Output (MIMO) systems realizing directive beamforming require reliable estimation of the wireless propagation channel. However, mmWave channels are characterized by high variability that severely challenges their recovery over short training periods. Current channel estimation techniques exploit either the channel sparsity in the beamspace domain or its low rank property in the antenna domain, nevertheless, they still require large numbers of training symbols for satisfactory performance. In this paper, we present a novel channel estimation algorithm that jointly exploits the latter two properties of mmWave channels to provide more accurate recovery, especially for shorter training intervals. The proposed iterative algorithm is based on the Alternating Direction Method of Multipliers (ADMM) and provides the global optimum solution to the considered convex mmWave channel estimation problem with fast convergence properties.  
\end{abstract}


%
\section{Introduction}
Near-optimal BeamForming (BF) performance in millimeter Wave (mmWave) massive Multiple Input Multiple Output (MIMO) systems employing Hybrid analog/digital BF (HBF) architectures necessitates reliable Channel State Information (CSI) knowledge. This knowledge is however very challenging to acquire in practice due to the very large numbers of transceiver antenna elements and the high channel variability \cite{Heath16}. Several approaches requiring receiver feedback have been lately proposed for designing BF vectors adequate for CSI estimation \cite{Venugopal17, Alkhateeb_JSTSP_2014}. On the other side, static dictionaries or beam training techniques without receiver feedback have also been adopted for beam codebook designs \cite{ Mendez_Access_2016, Lee16, GCA_2016}. In these studies, CSI estimation has been treated as a compressive sensing problem \cite{Donoho2006}, where the Orthogonal Matching Pursuit (OMP) algorithm \cite{5895106} has usually been adopted to recover the sparse channel gain vector. However, the performance of the aforementioned channel estimation techniques is usually limited by the codebook design, since beam dictionaries suffer from power leakage due to the discretization of the angles of arrival (AoA) and departure. Very recently in \cite{8122055}, mmWave CSI estimation that exploits both the sparsity and low rank properties of mmWave MIMO channels via a two independent stages procedure (one stage per each property) was proposed.

In this paper, we present a novel \textit{joint optimization formulation} for mmWave massive MIMO channel estimation incorporating both the sparsity and low rank properties, which possesses a global optimum solution due to its convexity property. To achieve the optimum solution, we capitalize on a recently developed theory of matrix completion with side information \cite{Lu16}, which we deploy together with the channel's beamspace representation. We develop an algorithm based on the Alternating Direction Method of Multipliers (ADMM) \cite{Boyd11} for efficient recovery of massive MIMO channel matrices. It is shown through representative simulation results that the proposed algorithm exhibits faster convergence and improved performance in terms of Mean Squared Error (MSE) for channel estimation with short training length, when compared with other state-of-the-art techniques \cite{Mendez_Access_2016, 8122055, 7869633, Cai2010}.

\subsubsection*{Notation} Fonts $\alpha$, $\mathbf{a}$, and $\mathbf{A}$ denote a scalar, a vector, and a matrix, respectively. $\mathbf{A}^T$, $\mathbf{A}^*$, $\mathbf{A}^H$, and $\Vert \mathbf{A}\Vert_F$ represent $\mathbf{A}$'s transpose, conjugate transpose, Hermitian transpose, and Frobenius norm. Operands $\circ$ and $\otimes$ denote the matrix Hadamard and Kronecker products, respectively, $\rm vec(\cdot)$ concatenates the columns of a matrix into a vector, and $\rm unvec(\cdot)$ is the inverse operation; $\Vert \mathbf{A} \Vert_* \triangleq \sum_{i=1}^r \sigma_i$ is the nuclear norm with $\sigma_i$'s being the $r$ singular values of $\mathbf{A}$; $\Vert \mathbf{A} \Vert_1 \triangleq \mathrm{max}_{1\leq j\leq N} \sum_{i=1}^M \vert [\mathbf{A}]_{ij} \vert $ ($\mathbf{A}\in\mathbb{C}^{M \times N}$) with $[\mathbf{A}]_{ij}$ denoting $\mathbf{A}$'s $(i,j)$-th element; $\mathcal{E}\{\cdot \}$ is the expected value. $\mathbf{A}\in\{0, 1\}^{M\times N}$ implies that $\mathbf{A}$'s elements are taken independently and with equal probability from the binary set $\{0,1\}$.

\section{System and Channel Models}\label{sec:Section_2}
We consider a $N_{\rm R} \times N_{\rm T}$ massive MIMO system operating over quasi-static mmWave channels, and adopting analog BF with switches \cite{Mendez_Access_2016} for the purpose of channel estimation. This cost and energy efficient BF scheme, which is sufficient for the channel estimation presented in this paper, can be realized with any available HBF architectures \cite{Molisch_HBF_2017}. Assuming that the channel $\mathbf{H}\in\mathbb{C}^{N_{\rm R} \times N_{\rm T}}$ remains static during the transmission of $T$ unit power training symbols $s[t] \in \mathbb{C}$, $\forall$$t=1,2,\ldots,T$, the post-processed received signal at the $N_{\rm R}$-element Receiver (RX) is expressed as $r[t]\triangleq\sqrt{P_t}\mathbf{w}^T\mathbf{H}\mathbf{f}s[t]+n[t]$, where $P_t$ is the Transmitter (TX) power, $\mathbf{w}\in\{0, 1\}^{N_{\rm R}}$ and $\mathbf{f}\in\{0, 1\}^{N_{\rm T}}$ denote the RX combining and TX precoding vectors, respectively, and $n[t]$ represents the zero-mean complex Additive White Gaussian Noise (AWGN) with variance $\sigma_n^2$. 

We adopt the geometric representation \cite{8122055, Alkhateeb_JSTSP_2014} for the mmWave MIMO channel, according to which $\mathbf{H}$ is given by 
\begin{equation}\label{eq:channel_model}
\mathbf{H} \triangleq \sum_{k=1}^{N_p} \alpha_k \mathbf{a}_{\rm R}(\phi_{\rm R}^{(k)}, \theta_{\rm R}^{(k)})  \mathbf{a}_{\rm T}^H(\phi_{\rm T}^{(k)}, \theta_{\rm T}^{(k)}),
\end{equation}
where $N_p$ denotes the number of propagation paths and $\alpha_k$ is the gain of the $k$-th path drawn from the complex Gaussian distribution $\mathcal{C}\mathcal{N}(0,1/2)$. $\mathbf{a}_{\rm T}^H(\phi_{\rm T}^{(k)},\theta_{\rm T}^{(k)})\in\mathbb{C}^{N_{\rm T}}$ and $\mathbf{a}_{\rm R}(\phi_{\rm R}^{(k)},\theta_{\rm R}^{(k)})\in\mathbb{C}^{N_{\rm R}}$ represent the TX and RX array response vectors, respectively, which are expressed as described in \cite[Sec. II.C]{Heath16} for uniform arrays. $\phi_{\rm T}^{(k)}$, $\theta_{\rm T}^{(k)}$ and $\phi_{\rm R}^{(k)}$, $\theta_{\rm R}^{(k)}$ are the physical elevation and azimuth angles of departure and arrival, respectively, which are generated according to the Laplace distribution \cite{4277071}. An alternative representation for $\mathbf{H}$ is based on the beamspace model \cite{1033686, Sayeed_TAP_2013} that is defined as 
\begin{equation}\label{eq:beamspace}
\mathbf{H} = \mathbf{D}_{\rm R}\mathbf{Z}\mathbf{D}_{\rm T}^H,
\end{equation} 
where $\mathbf{D}_{\rm R}\in\mathbb{C}^{N_{\rm R}\times N_{\rm R}}$ and $\mathbf{D}_{\rm T}\in\mathbb{C}^{N_{\rm T}\times N_{\rm T}}$ are unitary matrices based on the Discrete Fourier Transform (DFT). For Uniform Linear Arrays (ULAs), $\mathbf{D}_{\rm R}$ and $\mathbf{D}_{\rm T}$ are the normalized DFT matrices, whereas for the planar case they are the normalized Khatri-Rao products of DFT matrices \cite{1033686}. It holds for both cases that $\mathbf{D}_{\rm R}^H \mathbf{D}_{\rm R} = \mathbf{I}_{N_{\rm R}}$ and $\mathbf{D}_{\rm T}^H \mathbf{D}_{\rm T} = \mathbf{I}_{N_{\rm T}}$ with $\mathbf{I}_N$ being the $N\times N$ identity matrix. Also in \eqref{eq:beamspace}, $\mathbf{Z}\in\mathbb{C}^{N_{\rm R}\times N_{\rm T}}$ contains only a few virtual channel gains with high amplitude, i.e., it is a sparse (or compressible) matrix.

\section{Proposed mmWave MIMO Channel Estimation}
\subsection{Problem Formulation}
Matrix completion \cite{Cai2010} for the recovery of the unknown elements of a matrix $\mathbf{H}$ has been recently extended to incorporate side knowledge about the structure or properties of $\mathbf{H}$ \cite{NIPS2015_5940}. Motivated by this idea, we consider the beamspace representation of $\mathbf{H}$ given by \eqref{eq:beamspace} as its side information; particularly, we assume that the unknown $\mathbf{H}$ is decomposed as $\mathbf{D}_{\rm R}\mathbf{S}\mathbf{D}_{\rm T}^H$ with $\mathbf{S}$ being the unknown matrix. On this premise, we formulate the following constrained Optimization Problem (OP) for the joint recovery of the unknown CSI matrix $\mathbf{H}$ and its beamspace representation via the unknown sparse channel gain matrix $\mathbf{S}$:
\begin{align}\label{eq:matrix_completion_opt_side_information}
\min_{\mathbf{H}, \mathbf{S}} &\,\, \tau_H \Vert \mathbf{H} \Vert_* + \tau_S \Vert \mathbf{S} \Vert_1
\nonumber \\  \text{subject to} &\,\,\mathbf{\Omega} \circ \mathbf{H} = \mathbf{H}_{\Omega} \textrm{ and } \mathbf{H} = \mathbf{D}_{\rm R} \mathbf{S}  \mathbf{D}_{\rm T}^H,
\end{align} 
where $\mathbf{H}$'s nuclear norm in the objective imposes its low rank property, whereas the $\ell_1$-norm of $\mathbf{S}$ enforces its sparse structure. Also, constraint $\mathbf{H}=\mathbf{D}_{\rm R}\mathbf{S}\mathbf{D}_{\rm T}^H$ refers to $\mathbf{H}$'s representation given by \eqref{eq:beamspace}, and the weighting factors $\tau_H, \tau_S>0$ depend in general on the number of the mmWave channel propagation paths.

Matrix $\mathbf{\Omega}\in\{0,1\}^{N_{\rm R}\times N_{\rm T}}$ is composed of $M$ ones and $N_{\rm R}N_{\rm T}-M$ zeros, hence $\Vert \mathbf{\Omega} \Vert_0 = M$. The positions of its unity elements are randomly chosen in a uniform fashion over the set $\Omega\triangleq\{1,2,\ldots,N_{\rm R}N_{\rm T}\}$ \cite{Cai2010, NIPS2015_5940}. The matrix $\mathbf{H}_{\Omega}$ represents the subsampled estimated channel matrix and contains $M$ non-zero entries following the same pattern with $\mathbf{\Omega}$. These entries are derived prior to the solution of \eqref{eq:matrix_completion_opt_side_information}, based on the training procedure which is described in the next subsection. Clearly, $\mathbf{H}$'s estimation error from \eqref{eq:matrix_completion_opt_side_information} depends on the value of $M$ ($0\leq M\leq N_{\rm R}N_{\rm T}$) and the estimation accuracy of $\mathbf{H}_{\Omega}$'s elements. Note also that \eqref{eq:beamspace} may introduce additional errors due to the angle discretization effect \cite{Heath16, Alkhateeb_JSTSP_2014}.

\subsection{Proposed ADMM-based Solution}
The OP of \eqref{eq:matrix_completion_opt_side_information} is a two-objective convex problem, and thus, it possesses a global optimum which can be efficiently found via alternating optimization techniques \cite{Boyd11}. We first introduce the two auxiliary matrix variables $\mathbf{Y}\in\mathbb{C}^{N_{\rm R} \times N_{\rm T}}$ and $\mathbf{C}\triangleq\mathbf{Y}-\mathbf{D}_{\rm R}\mathbf{S} \mathbf{D}_{\rm T}^H$ to reformulate the targeted OP in the following equivalent form:
\begin{align}\label{eq:mc_opt_side_information_splitted}
\min_{\mathbf{H}, \mathbf{Y}, \mathbf{S}, \mathbf{C}} &\tau_H \Vert \mathbf{H} \Vert_* + \tau_S \Vert \mathbf{S} \Vert_1 + \frac{1}{2} \Vert \mathbf{C} \Vert_F^2 + \frac{1}{2} \Vert \boldsymbol{\Omega} \circ \mathbf{Y} - \mathbf{H}_{\Omega} \Vert_F^2  \nonumber 
\\ \text{subject to}& \,\, \mathbf{H} = \mathbf{Y} \text{ and }  \mathbf{C} = \mathbf{Y} - \mathbf{D}_{\rm R} \mathbf{S}  \mathbf{D}_{\rm T}^H.
\end{align}
Note that now the third and fourth terms in the objective take into account possible noise on $\mathbf{H}_{\Omega}$ in the problem formulation. Although OP in \eqref{eq:mc_opt_side_information_splitted} seems more complex than that in \eqref{eq:matrix_completion_opt_side_information}, it has separate blocks of variables (i$.$e$.$, a separable cost function). This property enables ADMM utilization, and consequently, the augmented Lagrangian function of \eqref{eq:mc_opt_side_information_splitted} is given by
\begin{align*} 
&\mathcal{L}_1 \big(\mathbf{H}, \mathbf{Y}, \mathbf{S}, \mathbf{C}, \mathbf{Z}_1, \mathbf{Z}_2 \big) \triangleq \tau_H \Vert \mathbf{H} \Vert_* + \tau_S \Vert \mathbf{S} \Vert_1 + \frac{1}{2} \Vert \mathbf{C} \Vert_F^2 \nonumber
\\&+ \frac{1}{2} \Vert \boldsymbol{\Omega} \circ \mathbf{Y} - \mathbf{H}_{\Omega} \Vert_F^2 + \text{tr}\big(\mathbf{Z}_1^H (\mathbf{H}- \mathbf{Y}) \big) + \frac{\rho}{2} \Vert \mathbf{H}-\mathbf{Y} \Vert_F^2 \nonumber
\\& + \text{tr}\big(\mathbf{Z}_2^H (\mathbf{Y}-\mathbf{D}_{\rm R} \mathbf{S} \mathbf{D}_{\rm T}^H - \mathbf{C}) \big) + \frac{\rho}{2} \Vert \mathbf{Y}-\mathbf{D}_{\rm R} \mathbf{S} \mathbf{D}_{\rm T}^H - \mathbf{C} \Vert_F^2,
\end{align*}
where $\mathbf{Z}_1,\mathbf{Z}_2\in\mathbb{C}^{N_{\rm R}\times N_{\rm T}}$ are dual variables (the Lagrange multipliers) adding the constraints of \eqref{eq:mc_opt_side_information_splitted} to the cost function, and $\rho$ denotes ADMM's stepsize. According to standard ADMM, at the $\ell$-th algorithmic iteration with $\ell=0,1,\ldots$ the following separate sub-problems need to be solved:
\begin{align}
\mathbf{H}^{(\ell+1)}\!\!&=\!\text{arg} \min_{\mathbf{H}} \mathcal{L}_1\big(\mathbf{H}, \mathbf{Y}^{(\ell)}, \mathbf{S}^{(\ell)}, \mathbf{C}^{(\ell)}, \mathbf{Z}_1^{(\ell)}, \mathbf{Z}_2^{(\ell)} \big), \label{eq:admm_si_1}\\
\mathbf{Y}^{(\ell+1)}\!\!&=\!\text{arg} \min_{\mathbf{Y}} \mathcal{L}_1\big(\mathbf{H}^{(\ell+1)}, \mathbf{Y}, \mathbf{S}^{(\ell)}, \mathbf{C}^{(\ell)}, \mathbf{Z}_1^{(\ell)}, \mathbf{Z}_2^{(\ell)} \big), \label{eq:admm_si_2}\\
\mathbf{S}^{(\ell+1)}\!\!&=\!\text{arg} \min_{\mathbf{S}} \mathcal{L}_1\big(\mathbf{H}^{(\ell+1)}, \mathbf{Y}^{(\ell+1)}, \mathbf{S}, \mathbf{C}^{(\ell)}, \mathbf{Z}_1^{(\ell)}, \mathbf{Z}_2^{(\ell)} \big), \label{eq:admm_si_3}\\
\mathbf{C}^{(\ell+1)}\!\!&=\!\text{arg} \min_{\mathbf{C}} \mathcal{L}_1\big(\mathbf{H}^{(\ell+1)}, \mathbf{Y}^{(\ell+1)}, \mathbf{S}^{(\ell+1)}, \mathbf{C}, \mathbf{Z}_1^{(\ell)}, \mathbf{Z}_2^{(\ell)} \big), \label{eq:admm_si_4}\\
\mathbf{Z}_1^{(\ell+1)}\!\!&=\!\mathbf{Z}_1^{(\ell)} + \rho \big( \mathbf{H}^{(\ell+1)}\!-\!\mathbf{Y}^{(\ell+1)}\big), \label{eq:admm_si_5} \\
\mathbf{Z}_2^{(\ell+1)}\!\!&=\!\mathbf{Z}_2^{(\ell)} + \rho \big( \mathbf{Y}^{(\ell+1)}\!-\!\mathbf{D}_{\rm R} \mathbf{S}^{(\ell+1)} \mathbf{D}_{\rm T}^H\!-\!\mathbf{C}^{(\ell+1)} \big). \label{eq:admm_si_6}
\end{align}
Note that for the initialization $\ell=0$: $\mathbf{H}^{(0)}=\mathbf{Z}_1^{(0)}=\mathbf{Z}_2^{(0)}=\mathbf{0}$.

To proceed with the formulation of the proposed algorithm, we derive closed form solutions for the problems $\eqref{eq:admm_si_1}-$\eqref{eq:admm_si_4}. First, to solve \eqref{eq:admm_si_1}, we reformulate $\mathcal{L}_1$ to $\mathcal{L}_2$ as follows, where the terms not affecting the minimization over $\mathbf{H}$ were removed and the term $\Vert \frac{1}{\rho} \mathbf{Z}_1^{(\ell-1)} \Vert_F^2$ was added:
\begin{equation}\label{eq:L_2}
\mathcal{L}_2(\mathbf{H}) \triangleq \tau_H \Vert \mathbf{H} \Vert_* + \frac{\rho}{2} \Vert  \mathbf{H} - ( \mathbf{Y}^{(\ell)} - \frac{1}{\rho} \mathbf{Z}_1^{(\ell)}) \Vert_F^2.
\end{equation}
Given the Lagrangian in \eqref{eq:L_2}, the solution of \eqref{eq:admm_si_1} is obtained from the Singular Value Thresholding (SVT) operator \cite{Cai2010}: 
\begin{equation}\label{eq:admm_x}
\mathbf{H}^{(\ell+1)} = \mathbf{U} \textrm{diag} \big(\{ \textrm{sign}(\zeta_i)\max(\zeta_i, 0) \}_{1\le i\le r}\big) \mathbf{V}^H,
\end{equation}
where $\mathbf{U} \in \mathbb{C}^{N_{\rm R} \times r}$ and $\mathbf{V} \in \mathbb{C}^{N_{\rm R} \times r}$ are the left and right singular vector matrices of the matrix $(\mathbf{Y}^{(\ell)} - \frac{1}{\rho} \mathbf{Z}_1^{(\ell)})$, respectively, and $\zeta_i \triangleq \sigma_i - \tau/\rho$ with $\sigma_i$'s denote its $r$ singular values. Similarly, to derive the solution of \eqref{eq:admm_si_2}, we reformulate $\mathcal{L}_1$ to the following Lagrangian function $\mathcal{L}_3$ of $\mathbf{Y}$:
\begin{align}
&\mathcal{L}_3(\mathbf{Y}) \triangleq \Vert \mathbf{\Omega}  \circ \mathbf{Y} - \mathbf{H}_{\Omega} \Vert_F^2 + \frac{\rho}{2} \Vert \frac{1}{\rho} \mathbf{Z}_1^{(\ell)}  + \mathbf{H}^{(\ell+1)} - \mathbf{Y} \Vert_F^2 \nonumber \\ &+  \frac{\rho}{2} \Vert \frac{1}{\rho} \mathbf{Z}_2^{(\ell)} + \mathbf{C}^{(\ell)} - \mathbf{Y} + \mathbf{D}_{\rm R} \mathbf{S}^{(\ell)} \mathbf{D}_{\rm T}^H \Vert_F^2,
\end{align}
which can be equivalently expressed based on the Krockecker vectorization and the Hadamard element-wise property as
\begin{align}\label{eq:L_3}
\mathcal{L}_3(\mathbf{y}) = &\frac{1}{2} \Vert \mathbf{A} \mathbf{y} - \mathbf{h}_{\Omega} \Vert_2^2 + \frac{\rho}{2}  \Vert \frac{1}{\rho} \mathbf{z}_1^{(\ell)}  + \mathbf{h}^{(\ell+1)} - \mathbf{y} \Vert_2 ^2 \nonumber \\ &+  \frac{\rho}{2} \Vert \frac{1}{\rho} \mathbf{z}_2^{(\ell)} + \mathbf{c}^{(\ell)} - \mathbf{y} + \mathbf{B} \mathbf{s}^{(\ell)} \Vert_2^2.
\end{align}
In \eqref{eq:L_3}, $\mathbf{B} \triangleq \mathbf{D}_{\rm T}^* \otimes \mathbf{D}_{\rm R}$ and $\mathbf{A} \triangleq \sum_{i=1}^{N_{\rm R}} \textrm{diag}([\mathbf{\Omega}]_i)^T \otimes \mathbf{E}_{ii}$ where $[\mathbf{\Omega}]_i$ denoting $\mathbf{\Omega}$'s $i$-th row and $\mathbf{E}_{ii}$ obtained from the $N_{\rm R}\times N_{\rm R}$ all-zero matrix after inserting a unity value at its $(i,i)$-th position. Also, small boldfaced letters are the $\rm vec(\cdot)$ results of their capital equivalents. Then, \eqref{eq:L_3} for \eqref{eq:admm_si_2} is minimized with:
\begin{align}\label{eq:admm_y}
\mathbf{y}^{(\ell+1)} = &(\mathbf{A}^H \mathbf{A} + 2 \rho \mathbf{I})^{-1} (\mathbf{z}_1^{(\ell)} + \rho \mathbf{h}^{(\ell+1)} \nonumber \\ &+ \mathbf{A}^H \mathbf{h}_{\Omega} + \mathbf{z}_2^{(\ell)} + \rho \mathbf{c}^{(\ell)} + \rho \mathbf{B}\mathbf{s}^{(\ell)}),
\end{align}
which is finally used to obtain $\mathbf{Y}^{(\ell+1)} = \mathrm{unvec}(\mathbf{y}^{(\ell+1)})$.

To find $\mathbf{S}$ solving \eqref{eq:admm_si_3}, we reformulate $\mathcal{L}_1$ to $\mathcal{L}_4$ as follows:
\begin{equation}
\mathcal{L}_4(\mathbf{S}) \triangleq \tau_S \Vert \mathbf{S} \Vert_1 + \frac{\rho}{2} \Vert \mathbf{D}_{\rm R}^H (\frac{1}{\rho} \mathbf{Z}_2^{(\ell)} - \mathbf{C}^{(\ell)} + \mathbf{Y}^{(\ell+1)})\mathbf{D}_{\rm T} + \mathbf{S} \Vert_F^2,
\end{equation}
where we have used the property that $\mathbf{D}_{\rm T}$ and $\mathbf{D}_{\rm R}$ are unitary matrices. By performing vectorization, $\mathcal{L}_4$ is equivalent to a standard LASSO problem \cite{Tibshirani94regressionshrinkage}, namely
\begin{equation} \label{eq:admm_si_3_lasso}
\mathrm{arg} \min_{\mathbf{s}} \tau_S \Vert \mathbf{s} \Vert_1 + \frac{\rho}{2} \Vert \mathbf{s} - \mathbf{v}^{(\ell+1)} \Vert_2^2,
\end{equation}
where $\mathbf{s}^{(\ell)} \triangleq \mathrm{vec}(\mathbf{S}^{(\ell)})$ and $\mathbf{v}^{(\ell+1)} \triangleq \mathrm{vec}(\mathbf{V}^{(\ell+1)})$ with
\begin{equation}\label{eq:admm_si_v2}
\mathbf{V}^{(\ell+1)} \triangleq\mathbf{D}_{\rm R}^H (\frac{1}{\rho} \mathbf{Z}_2^{(\ell)} - \mathbf{C}^{(\ell)} + \mathbf{Y}^{(\ell+1)})\mathbf{D}_{\rm T}.
\end{equation}
The solution of \eqref{eq:admm_si_3_lasso} is thus given by
\begin{align}\label{eq:s_l}
\mathbf{s}^{(\ell+1)}\!\!&=\!\mathrm{sign}(\mathtt{Re}(\mathbf{v}^{(\ell+1)})) \circ \max \big(\vert \mathtt{Re}(\mathbf{v}^{(\ell+1)}) \vert - \tau_S', 0 \big) \nonumber \\
&+ \mathrm{sign}(\mathtt{Im}(\mathbf{v}^{(\ell+1)})) \circ \max \big(\vert \mathtt{Im}(\mathbf{v}^{(\ell+1)}) \vert - \tau_S',0 \big),
\end{align}
where $\tau_S'\triangleq\tau_S/\rho$, and the $\mathrm{max}(\cdot)$ and the sign operator $\mathrm{sign}(\cdot)$ are applied component wise. The resulting vector in \eqref{eq:s_l} is then transformed into matrix form as $\mathbf{S}^{(\ell+1)} = \mathrm{unvec}(\mathbf{s}^{(\ell+1)})$.

To finally solve \eqref{eq:admm_si_4} for $\mathbf{C}$, we reformulate $\mathcal{L}_1$ as follows:
\begin{equation*}
\mathcal{L}_5(\mathbf{C})\!\!=\!\frac{1}{2} \Vert \mathbf{C} \Vert_F^2 + \frac{\rho}{2} \Vert \frac{1}{\rho} \mathbf{Z}_2^{(\ell)} + \mathbf{Y}^{(\ell+1)}-\mathbf{D}_{\rm R} \mathbf{S}^{(\ell+1)} \mathbf{D}_{\rm T}^H - \mathbf{C} \Vert_F^2,
\end{equation*}
which is strictly convex with respect to $\mathbf{C}$. Taking the derivative and equating it to zero yields the closed form solution:
\begin{equation}\label{eq:admm_update_c}
\mathbf{C}^{(\ell+1)}\!\!=\!\frac{\rho}{\rho+1} \big(\mathbf{Y}^{(\ell+1)} - \mathbf{D}_{\rm R}\mathbf{S}^{(\ell+1)}\mathbf{D}_{\rm T}^H + \frac{1}{\rho} \mathbf{Z}_2^{(\ell)} \big).
\end{equation}
Expressions \eqref{eq:admm_si_5} and \eqref{eq:admm_si_6} including the dual variable updates can be straightforwardly computed using \eqref{eq:admm_y}, \eqref{eq:s_l}, and \eqref{eq:admm_update_c}.

The previously described ADMM steps constituting the proposed mmWave massive MIMO CSI estimation technique are summarized in Algorithm~\ref{algorithm:admm_si}. After a predefined number of algorithmic iterations $I_{\text{max}}$, the output of this algorithm is the estimated MIMO channel matrix $\mathbf{H}^{(I_{\mathrm{max}})}$.

\subsubsection*{Computational complexity} The computational complexity of Algorithm~\ref{algorithm:admm_si} depends on the numbers of TX and RX antennas $N_{\rm T}$ and $N_{\rm R}$, as well as the number of iterations $I_{\text{max}}$. The most demanding step is in line $2$  with the update $\mathbf{H}^{(\ell+1)}$ which requires the computation of $\sigma_i$'s for the SVT operator. In general, the complexity of this computation for a $N_{\rm R} \times N_{\rm T}$ matrix is $\mathcal{O}(N_{\rm R}^2 N_{\rm T})$ \cite[Chapter 8.6]{Golub96}. However, the dominant singular values and vectors can be efficiently computed via incomplete singular value decomposition methods (e.g., Lanczos bidiagonalization algorithm \cite{Larsen98}) or via subspace tracking, thus the complexity can be further reduced to $\mathcal{O}(N_{\rm R}N_{\rm T})$.
\begin{algorithm}[t]
	\caption{ADMM-based MIMO Channel Estimation}
	\begin{algorithmic}[1]
		\REQUIRE $\mathbf{H}_{\Omega}$, $\mathbf{\Omega}$, $\mathbf{D}_{\rm R}$, $\mathbf{D}_{\rm T}$, $\rho$, $\tau_H$, $\tau_S$, and $I_{\max}$
		\ENSURE $\mathbf{H}^{(I_{\text{max}})}$ \\
		\textbf{Initialization:} $\mathbf{H}^{(0)} = \mathbf{S}^{(0)} = \mathbf{C}^{(0)} = \mathbf{Z}_2^{(0)} = \mathbf{Z}_1^{(0)} = \mathbf{0}$
		
		\FOR {$\ell=0,1,\ldots,I_{\max}-1$} 
		\STATE Update $\mathbf{H}^{(\ell+1)}$ using \eqref{eq:admm_x}.
		\STATE Update $\mathbf{Y}^{(\ell+1)}=\mathrm{unvec}(\mathbf{y}^{(\ell+1)})$ using \eqref{eq:admm_y}.
		\STATE Update $\mathbf{S}^{(\ell+1)} = \mathrm{unvec}(\mathbf{s}^{(\ell+1)})$ using \eqref{eq:s_l}.
		\STATE Update $\mathbf{C}^{(\ell+1)}$ using \eqref{eq:admm_update_c}.
		\STATE Update $\mathbf{Z}_1^{(\ell+1)}$ and $\mathbf{Z}_2^{(\ell+1)}$ using \eqref{eq:admm_si_5} and \eqref{eq:admm_si_6}.
		\ENDFOR
	\end{algorithmic}
	\label{algorithm:admm_si}	
\end{algorithm}

\subsubsection*{Channel Sub-Sampling $\mathbf{H}_{\Omega}$} Algorithm~\ref{algorithm:admm_si} will run at every channel coherence interval requiring as input the estimation of a sub-sampled version of $\mathbf{H}$. We adopt the training procedure described in Section~\ref{sec:Section_2} to estimate $M\ll N_{\rm R}N_{\rm T}$ non-zero elements of $\mathbf{H}_{\Omega}$ at respective training instances (i.e., $T=M$). Specifically, to estimate the $(i,j)$-th non-zero element of $\mathbf{H}_{\Omega}$ at the $t$-th training instance, we use the training symbol $s[t]$ and set $\mathbf{w}=\mathbf{e}_i$ and $\mathbf{f}=\mathbf{e}_j$ as the RX combining and TX precoding vectors, respectively, having zeros expect for their $i$-th and $j$-th positions, respectively, which contain ones. In fact, only one pair of TX and RX antennas is activated at each instance $t$. These training BF vectors can be efficiently realized with any available HBF architecture \cite{Molisch_HBF_2017} by including switches to the analog phase shifters. We note that in conventional estimation of sparse mmWave channels with analog BF training vectors implemented via antenna switches \cite{Mendez_Access_2016}, $\mathbf{w}\in\{0, 1\}^{N_{\rm R}}$ and $\mathbf{f}\in\{0, 1\}^{N_{\rm T}}$ are used at each training instance $t$. Then, all $T$ post-processed received signals are used for designing $\mathbf{H}$'s estimations. This training procedure is inherently different from the aforedescribed proposed one for estimating $\mathbf{H}_{\Omega}$.


\section{Simulation Results and Discussion}
We consider the examples of $32 \times 32$ and $64 \times 64$ MIMO systems equipped with ULAs at both TX and RX sides and operating over a 90GHz mmWave channel. The azimuth angles of arrival and departure have been generated from the Laplace distribution with standard deviation $50^\circ$. As benchmark CSI estimation techniques we have considered: 1) OMP \cite{Mendez_Access_2016}; 2) Vector Approximate Message Passing (VAMP) \cite{7869633}; 3) SVT \cite{Cai2010}; and the 4) Two-Stage estimation exploiting both Sparsity and low Rankness (TSSR) \cite{8122055}. Note that OMP and VAMP exploit only the sparsity of the channel matrix, while SVT capitalizes only on its low rank property. TSSR exploits both properties by first employing the SVT operator to recover the channel matrix, and then uses it as input to VAMP. The maximum number of iterations for SVT, VAMP, and the proposed algorithm was set to $I_{\textrm{max}}=100$, which from our experiments was verified as adequate for their convergence. For SVT we used $\tau=\rho \Vert \mathbf{H}_{\Omega} \Vert$ with $\rho=\frac{3 M}{(N_{\rm R} N_{\rm T})}$, for OMP and VAMP algorithms the parameter for channel sparsity was set to $N_p$, and we have considered $\tau_H=\rho \Vert \mathbf{H}_{\Omega} \Vert$ with $\rho=0.005$ and $\tau_S=\frac{0.1}{(1-10\log(\sigma_n^2))}$ for the proposed Algorithm~\ref{algorithm:admm_si}.
\begin{figure}[t]
	\centering
	\includegraphics[scale=0.49]{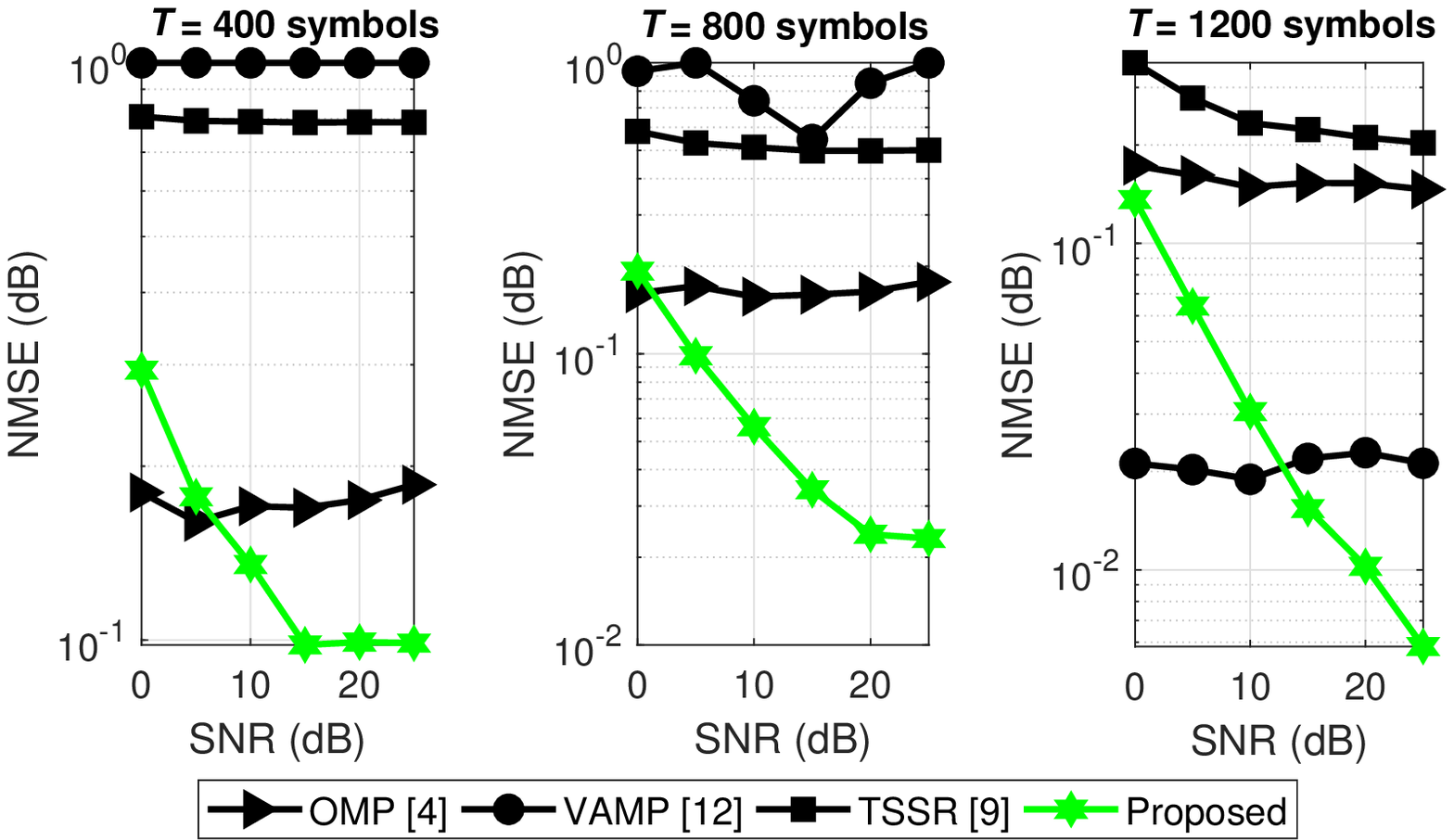}
	\vspace{-1.5em}
	\caption{NMSE w.r.t. transmit SNR for a $64\times64$ MIMO channel with $N_p=2$ and different $T$ values.}
	\vspace{-0.5em}
	\label{fig:figure1}
\end{figure}
\begin{figure}[t]
	\centering
	\vspace{-0.5em}
	\includegraphics[scale=0.49]{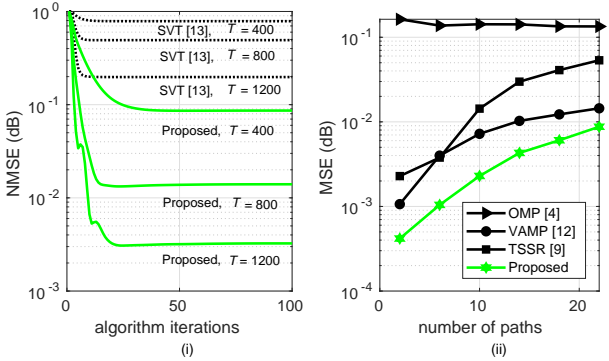}
	\vspace{-1.5em}
	\caption{NMSE for a $64\times64$ MIMO channel and $30$dB transmit SNR w.r.t. (i) algorithmic iterations and different $T$; and (ii) $N_p$ for $T=2000$.}
	\vspace{-0.8em}
	\label{fig:figure2}
\end{figure}

We compare the considered CSI estimation techniques both in terms of Normalized MSE (NMSE) performance and Achievable Spectral Efficiency (ASE) in bits/sec/Hz. For the SVT and proposed techniques that are based on matrix completion, the speed of convergence was also investigated for different values of the training instances $T$. Denoting by $\hat{\mathbf{H}}$ the estimation for the true channel $\mathbf{H}$ with any of the considered techniques, NMSE was numerically evaluated as follows:
\begin{equation}
{\rm NMSE} \triangleq \mathcal{E}\{ 10 \log_{10} \Vert \hat{\mathbf{H}}-\mathbf{H}\Vert^2_F / \Vert \mathbf{H}\Vert^2_F \}.
\end{equation}
Note that for $I_{\text{max}}$ algorithmic iterations for the proposed technique, $\hat{\mathbf{H}}$ is given by $\mathbf{H}^{(I_{\mathrm{max}})}$. In addition, we have computed the following lower bound for ASE \cite{1312613, 7078106}:
\begin{equation}
{\rm ASE}\!\triangleq\!\mathcal{E}\!\left\{\!\log_2\!\text{det}\!\left(\mathbf{I}_{N_{\rm R}}\!+\!(N_{\rm T} N_{\rm R} (\sigma_n^2\!+\! \textrm{NMSE}))^{-1}\mathbf{H} \mathbf{H}^H \right)\!\right\}\!.
\end{equation}
For both latter expressions, the expectations were obtained from averaging $100$ independent Monte Carlo realizations.   

It is demonstrated that the proposed technique outperforms OMP, VAMP, and TSSR in terms of NMSE performance for low training lengths $T$. As shown, OMP performance is not improved over the training length $T$ or SNR due to the discretization error of the AoA. Also, VAMP is incapable of recovering the $64\times64$ MIMO channel matrix for small numbers ($<800$) of training symbols. This happens because VAMP is based on the calculation of the statistical information of the sparse signal, which cannot be captured for small $T$. However, for $T\geq800$ and low-to-medium transmit Signal-to-Noise Ratio (SNR) ($10\log_{10}(P_t/\sigma_n^2)$ in dB) values ($<16$dB), VAMP provides improved NMSE compared to the proposed algorithm. This behavior is due to the different training procedures between these two techniques. Specifically, the proposed training lacks of array gain, and hence, $\mathbf{H}_{\Omega}$ estimation will be in general more noisy than channel estimation with VAMP. Nevertheless, this noisy estimation becomes less impactfull and less severe as transmit SNR increases. It is also evident in Fig$.$~\ref{fig:figure1} that TSSR, which is based on successive application of SVT and VAMP algorithms, cannot recover the channel for small $T$ values. This indicates that the independent treatment of each stage does not permit the joint exploitation of the channel sparsity and low rank properties.
\begin{figure}[t]
	\centering
	\includegraphics[scale=0.47]{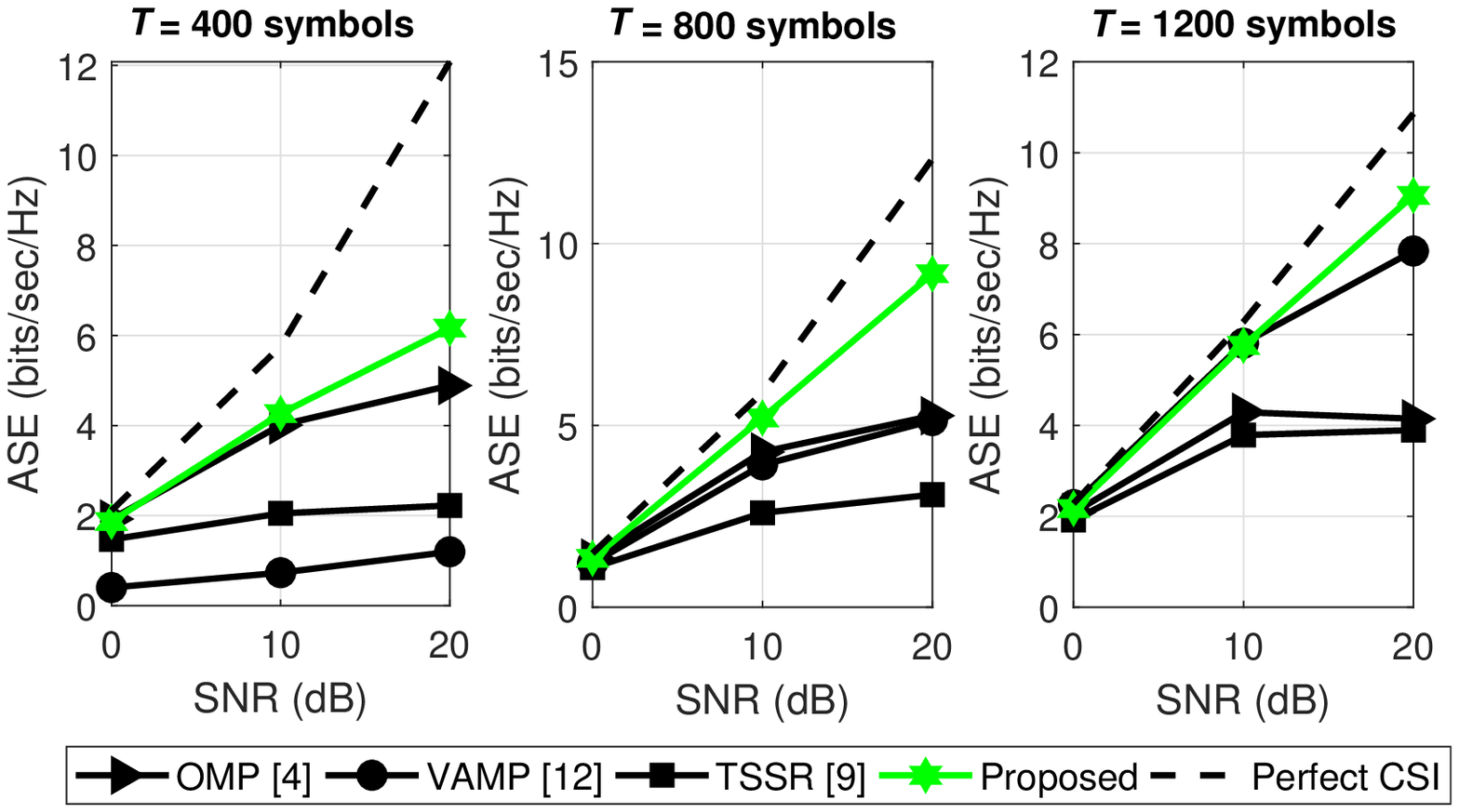}
	\vspace{-0.9em}
	\caption{ASE w.r.t. transmit SNR for a $32\times32$ MIMO channel with $N_p=2$ and different $T$ values.}
	\vspace{-1em}
	\label{fig:figure3}	
\end{figure}

The fast convergence of the proposed algorithm, even for very small training lengths $T$, is illustrated in Fig$.$~\ref{fig:figure2}(i), and compared with that of the standard SVT technique. As shown, the proposed algorithm converges to smaller NMSE performances for all $T$ values. In Fig$.$~\ref{fig:figure2}(ii), we depict the impact of the $N_p$ value (i$.$e$.$, the number of mmWave channel propagation paths) to NMSE of all considered estimation techniques. As expected, CSI estimation performance gets worse as $N_p$ increases, however, the proposed technique outperforms all others providing good NMSE for quite large $N_p$ values. We finally present ASE performance of all techniques for $N_p=2$ in Fig$.$~\ref{fig:figure3}. In this figure, ASE with perfect CSI is also sketched. Clearly, the proposed technique outperforms all others for all considered $T$ values. Interestingly, the higher NMSE of the proposed technique for low-to-mid SNR values and $T=1200$ shown in Fig$.$~\ref{fig:figure1} does not affect its superiority in terms of ASE.

Putting all above together, the proposed matrix completion estimation of mmWave massive MIMO channels leveraging jointly their sparsity and low rank properties outperforms the state-of-the-art techniques requiring short training periods. Our performance results showed that the convergence of the proposed ADMM-based iterative approach is relatively fast. 

\newpage
\bibliographystyle{IEEEtran}
\bibliography{IEEEabrv,literature}

\begin{thebibliography}{10}
\providecommand{\url}[1]{#1}
\csname url@samestyle\endcsname
\providecommand{\newblock}{\relax}
\providecommand{\bibinfo}[2]{#2}
\providecommand{\BIBentrySTDinterwordspacing}{\spaceskip=0pt\relax}
\providecommand{\BIBentryALTinterwordstretchfactor}{4}
\providecommand{\BIBentryALTinterwordspacing}{\spaceskip=\fontdimen2\font plus
\BIBentryALTinterwordstretchfactor\fontdimen3\font minus
  \fontdimen4\font\relax}
\providecommand{\BIBforeignlanguage}[2]{{%
\expandafter\ifx\csname l@#1\endcsname\relax
\typeout{** WARNING: IEEEtran.bst: No hyphenation pattern has been}%
\typeout{** loaded for the language `#1'. Using the pattern for}%
\typeout{** the default language instead.}%
\else
\language=\csname l@#1\endcsname
\fi
#2}}
\providecommand{\BIBdecl}{\relax}
\BIBdecl

\bibitem{Heath16}
R.~W. Heath, Jr., N.~Gonz{\'a}lez-Prelcic, S.~Rangan, W.~Roh, and A.~M. Sayeed,
  ``An overview of signal processing techniques for millimeter wave {MIMO}
  systems,'' \emph{IEEE J. Sel. Topics Signal Process}, vol.~10, no.~3, pp.
  436--453, Apr. 2016.

\bibitem{Venugopal17}
K.~Venugopal, A.~Alkhateeb, R.~W. Heath, Jr., and N.~Gonz{\'a}lez-Prelcic,
  ``Time-domain channel estimation for wideband millimeter wave systems with
  hybrid architecture,'' in \emph{Proc. IEEE ICASSP}, New Orleans, USA, Mar.
  2017, pp. 6493--6497.

\bibitem{Alkhateeb_JSTSP_2014}
A.~Alkhateeb, O.~E. Ayach, G.~Leus, and R.~W. Heath, Jr., ``Channel estimation
  and hybrid precoding for millimeter wave cellular systems,'' \emph{{IEEE}
  {J.} {S}el. {T}opics {S}ignal {P}rocess.}, vol.~8, no.~5, pp. 831--846, Oct.
  2014.

\bibitem{Mendez_Access_2016}
R.~M{\'e}ndez-Rial, C.~Rusu, N.~Gonz{\'a}lez-Prelcic, A.~Alkhateeb, and R.~W.
  Heath, Jr., ``Hybrid {MIMO} architectures for millimeter wave communications:
  {P}hase shifters or switches?'' \emph{{IEEE} {A}ccess}, vol.~4, pp. 247--267,
  Jan. 2016.

\bibitem{Lee16}
J.~Lee, G.~T. Gil, and Y.~H. Lee, ``Channel estimation via orthogonal matching
  pursuit for hybrid {MIMO} systems in millimeter wave communications,''
  \emph{IEEE Trans. Commun.}, vol.~64, no.~6, pp. 2370--2386, Jun. 2016.

\bibitem{GCA_2016}
G.~C. Alexandropoulos and S.~Chouvardas, ``Low complexity channel estimation
  for millimeter wave systems with hybrid {A/D} antenna processing,'' in
  \emph{Proc. IEEE GLOBECOM}, Washington D.C., USA, Dec. 2016, pp. 1--6.

\bibitem{Donoho2006}
D.~L. Donoho, ``Compressed sensing,'' \emph{IEEE Trans. Inf. Theory}, vol.~52,
  no.~4, pp. 1289--1306, Apr. 2006.

\bibitem{5895106}
T.~T. Cai and L.~Wang, ``Orthogonal matching pursuit for sparse signal recovery
  with noise,'' \emph{IEEE Trans. Inf. Theory}, vol.~57, no.~7, pp. 4680--4688,
  Jul. 2011.

\bibitem{8122055}
X.~Li, J.~Fang, H.~Li, and P.~Wang, ``Millimeter wave channel estimation via
  exploiting joint sparse and low-rank structures,'' \emph{IEEE Trans. Wireless
  Commun.}, vol.~17, no.~2, pp. 1123--1133, Feb. 2018.

\bibitem{Lu16}
J.~Lu, G.~Liang, J.~Sun, and J.~Bi, ``A sparse interactive model for matrix
  completion with side information,'' in \emph{Adv. Neural Inf. Process. Syst.
  29}, 2016, pp. 4071--4079.

\bibitem{Boyd11}
S.~Boyd, N.~Parikh, E.~C.~B. Peleato, and J.~Eckstein, ``Distributed
  optimization and statistical learning via the alternating direction method of
  multipliers,'' \emph{Found. Trends Mach. Learn.}, vol.~3, no.~1, pp. 1--122,
  Jan. 2011.

\bibitem{7869633}
P.~Schniter, S.~Rangan, and A.~K. Fletcher, ``Vector approximate message
  passing for the generalized linear model,'' in \emph{Proc. Asilomar CSSC},
  Pacific Grove, USA, Nov. 2016, pp. 1525--1529.

\bibitem{Cai2010}
J.~F. Cai, E.~J. Cand{\`e}s, and Z.~Shen, ``A singular value thresholding
  algorithm for matrix completion,'' \emph{SIAM J. Opt.}, vol.~20, no.~4, pp.
  1956--1982, 2010.

\bibitem{Molisch_HBF_2017}
A.~F. Molisch, V.~V. Ratnam, S.~Han, Z.~Li, S.~L.~H. Nguyen, L.~Li, and
  K.~Haneda, ``Hybrid beamforming for massive {MIMO}: {A} survey,''
  \emph{{IEEE} {Commun.} {M}ag.}, vol.~55, no.~9, pp. 134--141, Sep. 2017.

\bibitem{4277071}
A.~Forenza, D.~J. Love, and R.~W. Heath, Jr., ``Simplified spatial correlation
  models for clustered {MIMO} channels with different array configurations,''
  \emph{IEEE Trans. Veh. Technol.}, vol.~56, no.~4, pp. 1924--1934, Jul. 2007.

\bibitem{1033686}
A.~M. Sayeed, ``Deconstructing multiantenna fading channels,'' \emph{IEEE
  {Trans.} {S}ignal {P}rocess.}, vol.~50, no.~10, pp. 2563--2579, Oct. 2002.

\bibitem{Sayeed_TAP_2013}
J.~Brady, N.~Behdad, and A.~M. Sayeed, ``Beamspace {MIMO} for millimeter-wave
  communications: {S}ystem architecture, modeling, analysis, and
  measurements,'' \emph{{IEEE} {Trans.} {A}ntennas {P}ropag.}, vol.~61, no.~7,
  pp. 3814--3827, Oct. 2013.

\bibitem{NIPS2015_5940}
K.-Y. Chiang, C.-J. Hsieh, and I.~S. Dhillon, ``Matrix completion with noisy
  side information,'' \emph{Adv. Neural Inf. Process. Sys. 28}, pp. 3447--3455,
  2015.

\bibitem{Tibshirani94regressionshrinkage}
R.~Tibshirani, ``Regression shrinkage and selection via the {L}asso,'' \emph{J.
  Royal Stat. Society, Series B}, vol.~58, no.~1, pp. 267--288, 199.

\bibitem{Golub96}
G.~H. Golub and C.~F. Van~Loan, \emph{Matrix Computations (4th Ed.)}.\hskip 1em
  plus 0.5em minus 0.4em\relax Baltimore, MD, USA: Johns Hopkins University
  Press, 2013.

\bibitem{Larsen98}
R.~M. Larsen, ``Lanczos bidiagonalization with partial reorthogonalization,''
  1998.

\bibitem{1312613}
T.~Yoo and A.~Goldsmith, ``Capacity and power allocation for fading {MIMO}
  channels with channel estimation error,'' \emph{{IEEE} {Trans.} {I}nf.
  {T}heory}, vol.~52, no.~5, pp. 2203--2214, May 2006.

\bibitem{7078106}
L.~Berriche, K.~Abed-Meraim, and J.~C. Belfiore, ``Investigation of the channel
  estimation error on {MIMO} system performance,'' in \emph{Proc. European Sig.
  Proces. Conf.}, Antalya, Turkey, Sep. 2005, pp. 1--4.

\end{thebibliography}

\end{document}